\documentclass[11pt]{article}
\usepackage{amsfonts,amsmath,amssymb,mathrsfs,amsthm}

\usepackage{geometry}
\newtheorem{myDef}{Definition}[section]
\newtheorem{defn}[myDef]{Definition}
\newtheorem{prop}[myDef]{Proposition}
\newtheorem{exmp}[myDef]{Example}

\newtheorem{lem}[myDef]{Lemma}

\newtheorem{thm}[myDef]{Theorem}

\def\N{{\mathbb{N}}}
\def\Q{{\mathbb{Q}}}
\def\C{{\mathbb{C}}}

\def\Y{{\mathbb{Y}}}
\def\U{{\mathbb{U}}}
\def\W{{\mathbb{W}}}

\def\A{{\mathcal{A}}}
\def\F{{\mathcal{F}}}
\def\R{{\mathcal{R}}}
\def\CC{{\mathcal{C}}}

\def\grem{\hbox{\rm{grem}}}
\def\prem{\hbox{\rm{prem}}}
\def\lprem{\hbox{\rm{lprem}}}
\def\Res{\hbox{\rm{Res}}}
\def\sat{\hbox{\rm{sat}}}
\def\lsat{\hbox{\rm{lsat}}}
\def\dsat{\hbox{\rm{dsat}}}
\def\ldsat{\hbox{\rm{ldsat}}}
\def\lvar{\hbox{\rm{lv}}}
\def\init{\hbox{\rm{init}}}
\def\Zero{\hbox{\rm{Zero}}}
\def\lm{\hbox{\rm{lm}}}
\def\lc{\hbox{\rm{lc}}}
\def\lt{\hbox{\rm{lt}}}
\def\lcm{\hbox{\rm{lcm}}}
\def\val{\hbox{\rm{val}}}
\def\Val{\hbox{\rm{Val}}}
\def\Supp{\hbox{\rm{Supp}}}
\def\tr-{\hbox{\rm{tr-}}}
\def\gap{\hbox{\rm{gap}}}
\def\coeff{\hbox{\rm{lc}}}
\def\trop{\hbox{\rm{trop}}}

\def\ord{\hbox{\rm{ord}}}
\def\ld{\hbox{\rm{ld}}}
\def\sep{\hbox{\rm{sep}}}

\def\tdeg{\hbox{\rm{tdeg}}}

\begin{document}

\title{Tropical Differential  Gr\"obner Basis}

\author{Youren Hu and Xiao-Shan Gao\\
KLMM, UCAS, Academy of Mathematics and Systems Science\\
Chinese Academy of Sciences, Beijing 100190, China}
\date{}
\maketitle

\begin{abstract}
\noindent
In this paper, the tropical differential Gr\"obner basis is studied,
which is a natural generalization of the tropical Gr\"obner basis
to the recently introduced tropical differential algebra.
Like the differential Gr\"obner basis, the tropical differential Gr\"obner basis generally contains an infinite number of elements.
We give a Buchberger style criterion for the tropical differential Gr\"obner basis.
For ideals generated by homogeneous  linear differential polynomials with constant coefficients,
we give a complete algorithm to compute the tropical differential Gr\"obner basis.
\end{abstract}

\section{Introduction}

The Gr\"obner basis of a polynomial ideal is a basic tool in computational algebra,
which is designed over coefficient fields or rings with trivial valuations.
In \cite{JA}, Chan and Maclagan introduced an algorithm to
compute tropical Gr\"obner bases for polynomial ideals, where the coefficient
field has a non-trivial valuation.
Tropical Gro\"obner bases can be used to compute tropical varieties in
tropical algebraic geometry, when the base field has a nontrivial valuation,
such as $\Q_p$. Please refer to \cite{CTV,DM} for a detailed introduction
to tropical geometry.
In \cite{G}, algorithms to compute the tropical Gr\"{o}bner basis for modules over polynomial rings  were given.
In \cite{tgb1}, F5 style algorithms to compute tropical Gr\"obner bases were given.
In \cite{C}, it was shown that certain computation of tropical varieties
over fields with non-trivial valuations can be done with standard Gr\"obner basis in certain cases.

Grigoriev initiated the study of tropical differential algebra  by designing a polynomial-time algorithm to compute the solution of a system of tropical linear differential equations in one variable \cite{dg}.
In \cite{fund}, Aroca, Garay, and Toghani proved the fundamental theorem
for tropical differential algebraic geometry when the solutions under consideration
are univariate power series. More precisely, they proved that
$\trop(Sol(I))= Sol(\trop(I))$ for a differential ideal $I$,
which means that $I$ has a univariate power series with support $\phi$ as a solution
if and only if $\phi$ is a solution of the tropicalization of $I$.
As a consequence, a differential polynomial ideal $I$ has a power series solution with support $\overline{S}\in \mathcal{P}(\mathbb{N})^n$ if and only if the set of the corresponding initials $\{{\rm in}_{\overline{S}}(f):f\in I\}$ is monomial free\cite{fund}.
Therefore, the tropical differential Gr\"obner basis can be used to compute differential tropical varieties and to decide whether a differential polynomial ideal has a solution with a given support.
Similar to the algebraic case, tropical differential algebraic geometry
can be considered as the abstract generalization of the methods
of computing the power series solutions of differential equations
by comparing the terms with the lowest degree \cite{J2,J4,jssc-lcs}.

In this paper, we define the concept of tropical differential Gr\"obner basis
for differential polynomial ideals over differential fields with a
differential valuation and prove some of its basic properties,
which can also  be considered as the generalization of the
differential Gr\"obner basis \cite{GF,FO} to the tropical case.
Like the differential Gr\"obner basis \cite{GF,FO},
the tropical differential Gr\"obner basis generally contains an infinite number of elements.
We give a lower bound for the number of differentiations
in order to compute the tropical differential Gr\"obner basis
and give a partial Buchberger style algorithm.
For ideals generated by homogeneous linear differential polynomials with constant coefficients, we give a complete algorithm to compute the tropical differential Gr\"obner basis when the support is $s\N^n$ for $s\in\N$. The main technique contribution is to give an upper bound for the number of differentiations
in order to compute the finite tropical differential Gr\"obner basis.

The rest of this paper is organized as follows.
In section 2, we define the concept of tropical basis for a differential polynomial ideal.
In section 3, we define the concept of and tropical Gr\"obner basis for a differential polynomial ideal and give a partial Buchberger style algorithm.
In section 4, we give an algorithm for computing  a  tropical Gr\"obner basis of a differential ideal generated by homogeneous linear differential polynomials with constant coefficients.

\section{Differential tropic basis}
\label{sec:1}
Similar to the algebraic case, tropical differential Gr\"obner bases can be considered
as differential Gr\"obner bases over a differential field with valuations.

\begin{defn}[\cite{MR}]
Let $\mathcal{F}$ be an ordinary differential field of characteristic zero with a differential operator $\delta$, $\mathcal{C}$ its subfield of constants. A differential valuation of $\mathcal{F}$ is a valuation $\rm{v}$ of $\mathcal{F}$ that is trivial on $\mathcal{C}$
and such that $\mathcal{A}=\{x:\rm{v}(x)\geq 0\}$ and ${M}=\{x:\rm{v}(x)>0\}$ are respectively its valuation ring and maximal ideal, then $\mathcal{A}=\mathcal{C}+{M}$ and $a\in\mathcal{A}$, nonzero $b\in {M}$ implies $\delta(a)b/\delta(b)\in {M}$.
\end{defn}

Let $\R=\CC[[t]]$, where $\CC$ is an algebraically close field of characteristic zero and $\delta c=0$
for any $c\in \CC.$ We denote $\F=Frac(\R)$ the quotient field of $\R$.
For simplicity, we will work with the field $\F$.
{The results of this paper are correct over general differential fields with valuations.}


We define a mapping $\val:\F\setminus\{0\}\rightarrow \mathbb{N}$.
Let $\phi_i=\sum_{j\in\mathbb{N}}a_{ij}t^j\in \R$, $i=1,2$. The {\em support} of $\phi_i$ is defined to be  $\Supp(\phi_i):=\{j\in\mathbb{N}:a_{ij}\neq 0\}.$
Then $\val(\frac{\phi_1}{\phi_2})$ is defined to be $\min \Supp(\phi_1)-\min\Supp(\phi_2)$.
We add $\val(0)=+\infty.$
The mapping $\val$ has two simple properties: $\val(ab)=\val(a)+\val(b)$ and $\val(a+b)\geq \min\{\val(a),\val(b)\}.$
So $\val$ is a valuation of $\F$.

For $\phi=\frac{\phi_1}{\phi_2}\in\F,$ write $\phi$ as a Laurent series $\phi=c_{s}t^{s}+c_{s+1}t^{s+1}+\cdots$. Then we have $\val(\phi)=s.$
We denote $\bar{\phi}=c_{s}t^s$ and $\coeff(\phi)=c_{s}.$

\begin{lem}
$\val$ is a differential valuation of $\F$.
\end{lem}
{\noindent\bf Proof.}
It is easy to see that $\mathcal{A}=\{x:\val(x)\geq 0\}=\{\sum_{i=0}^\infty c_it^i:c_i\in \CC\}$
and ${M}=\{x:\val(x)>0\}=\{\sum_{i=1}^\infty c_it^i:c_i\in \CC\}$.
So $\mathcal{A}=\mathcal{C}+{M}$ holds.
Let $a=\sum_{i=0}^\infty a_it^i$ and $b=\sum_{j=d}^\infty b_dt^j$ with $d\geq 1$ and $b_d\neq 0.$
So $\val(a')\geq 0$, $\val(b)=d$ and $\val(b')=d-1$.
Then by the definition of $\val$, $\val(a'b/b)=\val(a')+\val(b)-\val(b')\geq 1$,
which implies that $a'b/b\in{M}$.
\qed

Let $\Y=\{y_1,y_2,\ldots,$ $y_n\}$ be a set of differential indeterminates
and $\mathcal{F}\{\Y\}$ the differential polynomial ring in $\Y$ over $\mathcal{F}$.
Set $\Theta\Y=\{\delta^j y_i:i=1,\ldots,n, j\in\mathbb{N}\}.$
For $f_1,\ldots,f_r$ in $\mathcal{F}\{\Y\}$, we denote
$(f_1,\ldots,f_r)$ and $[f_1,\ldots,f_r]$ to be the algebraic ideal
and the differential ideal generated by $f_1,\ldots,f_r$, respectively.
%
For $f\in \mathcal{F}\{\Y\}$, if $y_i$ appears in $f$,
the order of $f$ in $y_i$ is defined to be the maximal $j$ such that $\delta^jy_i$ occurs in $f$, denoted by $\ord(f,y_i)$.
If $y_i$ does not appear in $f$, set $\ord(f,y_i)=-\infty$.
The order of $f$ is defined as $\ord(f)=\max_{1\leq i\leq n}\{\ord(f,y_i)\}.$

A differential monomial in $\Y$ of order less than or equal to $r$ is an expression of the form
\begin{equation}\label{equ-1}
E_M:=\prod_{1\leq i\leq n,0\leq j\leq r}y_{ij}^{M_{ij}},
\end{equation}
where $y_{ij}=\delta^j(y_i)$ and $M=(M_{ij})_{1\leq i\leq n,0\leq j\leq r}\in\mathcal{M}_{n\times (r+1)}(\mathbb{N})$,
which is the set of all matrix  of size $n\times (r+1)$ with elements in $\mathbb{N}$.
The set of all differential monomials in $\Y$ is denoted by $\mathcal{M}_\Y.$
With the above notation, a differential polynomial $P\in\mathcal{F}\{\Y\}$ of order $r$
is a finite $\F$-linear combination of differential monomials with order $\le r$:
\begin{equation}\label{equ-2}
f=\sum_{M\in\mathcal{M}_{n\times (r+1)}(\mathbb{N}) }c_M E_M.
\end{equation}
We always assume $c_M\ne0$.

Denote $\mathcal{P}(\mathbb{N})$ to be the power set of $\mathbb{N}.$

\begin{defn}
Let $S\in\mathcal{P}(\mathbb{N})$. Define a mapping $\Val_S:\mathbb{N}\rightarrow \mathbb{N}\cup\{\infty\}$
as follows: $\Val_S(j):=s-j$ if $S\cap\mathbb{N}_{\geq j}\ne\varnothing$, where $s=\min\{\alpha\in S:\alpha\geq j\}$; $\Val_S(j):=\infty$ if $S\cap\mathbb{N}_{\geq j}=\varnothing.$
\end{defn}

\begin{defn}
Let ${\overline{S}}=(S_1,\ldots,S_n)\in \mathcal{P}(\mathbb{N})^n.$
We define a mapping $\Val_{{\overline{S}}}:\mathcal{M}_\Y\rightarrow \mathbb{N}\cup\{\infty\}$ given by
$\Val_{{\overline{S}}}(E_M)=\sum_{i=1}^n\sum_{j=0}^{r}M_{ij}\Val_{S_i}(j)$, where $E_M$ is of the form in (\ref{equ-1}).
\end{defn}

Let ${\overline{S}}=(S_1,\ldots,S_n)\in \mathcal{P}(\mathbb{N})^n$ and $f$ be of the form in (\ref{equ-2}).
Define $\Val_f=\min\{\val(c_M)+\Val_{{\overline{S}}}(E_M)\}.$
Then the {\em initial of $f$ with respect to ${\overline{S}}$} is
defined to be
\begin{equation}
{\rm in}_{{\overline{S}}}(f)=\sum_{\val(c_M)+\Val_{{\overline{S}}}(E_M)=\Val_f}
{{\overline{c_M}}}E_M\in \F\{\Y\}.
\end{equation}
Note that ${\rm in}_{{\overline{S}}}(f)=\infty$ if and only if $\Val_{{\overline{S}}}(E_M)=0$ for all $c_M\neq0.$

\begin{defn}
The {\em initial ideal} of a differential ideal $I\subset \F\{\Y\}$ with respect to ${\overline{S}}$ is
\begin{equation}\label{equ-3}
{\rm in}_{{\overline{S}}}(I)=(\\{\rm in}_{{\overline{S}}}(f):f\in I)\subset \F\{\Y\}.
\end{equation}
\end{defn}

Note that ${\rm in}_{{\overline{S}}}(I)$ is an algebraic ideal.
The following lemma gives the motivation for the definition.
Let ${\bar{y}}=(\overline{y_1},\cdots,\overline{y_n})\in \R^n.$ Denote $\trop({\bar{y}})=(\Supp(\overline{y_1}),\cdots,\Supp(\overline{y_n}))\in \mathcal{P}(\mathbb{N})^n.$
By \cite{fund}, ${\bar{y}}$ is a solution of $I$ if and only if $\{{\rm in}_{\trop({\bar{y}})}(f):f\in I\}$ contains no monomials.

\begin{lem}
Let ${\overline{S}}\in \mathcal{P}(\mathbb{N})^n.$
Then the set $\{{\rm in}_{{\overline{S}}}(f):f\in I\}$ contains no monomials if and only if ${\rm in}_{{\overline{S}}}(I)$ contains no monomials.
\end{lem}
{\noindent\bf Proof.}
The sufficiency is obvious.
It suffices to prove that if ${\rm in}_{{\overline{S}}}(I)$ has a monomial, then $\{{\rm in}_{{\overline{S}}}(f):f\in I\}$ has a monomial.
Suppose that $M=\sum_{i=1}^lg_i{\rm in}_{{\overline{S}}}(f_i)\in{\rm in}_{{\overline{S}}}(I)\cap{\mathcal M}_\Y$ with $\Val_{{\overline{S}}}(M)=a$, where $f_i\in I$ and $g_j\in\F\{\Y\}$.
If we write $g_i$ as $g_i=\sum_{j=0}^{l_i}g_{ij}$, where every term in $g_{ij}$ is of value $j$,
then $M$ can be represented as $M=\sum_{i=1}^l g_{i,a-\Val_{{\overline{S}}}(f_i)}{\rm in}_{{\overline{S}}}(f_i).$
So $f=\sum_{i=1}^l g_{i,a-\Val_{{\overline{S}}}(f_i)}f_i=
\sum_{i=1}^l g_{i,a-\Val_{{\overline{S}}}(f_i)}{\rm in}_{{\overline{S}}}(f_i)+
\sum_{i=1}^l g_{i,a-\Val_{{\overline{S}}}(f_i)}(f_i-{\rm in}_{{\overline{S}}}(f_i))$.
Since $\Val_{{\overline{S}}}(g_{i,a-\Val_{{\overline{S}}}(f_i)}(f_i-{\rm in}_{{\overline{S}}}(f_i)))>\Val_{{\overline{S}}}(g_{i,a-\Val_{{\overline{S}}}(f_i)}(f_i))=a
=\Val_{{\overline{S}}}(M)$,
we have ${\rm in}_{{\overline{S}}}(f)=M\in \{{\rm in}_{{\overline{S}}}(f):f\in I\}.$
\qed

\begin{defn}
For ${\overline{S}}\in \mathcal{P}(\mathbb{N})^n$ and a differential ideal $I\subset\F\{\Y\},$ a set $\mathcal{G}\subset I$ is called {\em tropical  basis} for $I$ with respect to ${\overline{S}}$
if ${\rm in}_{{\overline{S}}}(I)=({\rm in}_{{\overline{S}}}(\delta^ig):g\in \mathcal{G},i\in\mathbb{N})$.
\end{defn}

The concepts of tropical bases and tropical initials are quite different
from their algebraic and differential counterparts.
In the following, we use four examples to illustrate
sone of the key distinctions.

The following example shows that ${\rm in}_{{\overline{S}}}(\delta f)$
is generally not equal to $\delta ({\rm in}_{{\overline{S}}}(f))$.
\begin{exmp}
Let $S=\{0,2,4\}$ and  $f=yy''+y'\in\F\{y\}$. Then $\delta f=y'y''+yy'''+y''.$ We have ${\rm in}_{{S}}(\delta f)=y''$, ${\rm in}_S(f)=yy''$ and $\delta({\rm in}_{S}(f))=yy'''+y'y''$, which are not equal.
\end{exmp}

The following example shows that  $[{\rm in}_{{\overline{S}}}(f):f\in I]$
may contain more monomials than ${\rm in}_{{\overline{S}}}(I)$, which further justifies the
definition of ${\rm in}_{{\overline{S}}}(I)$.
\begin{exmp}
Let $f=ty''-3ty'+3y-3$, $I=[f]$, and $S=\{0,1,3\}$.
It is easy to verify that $\tilde{y}=at^2+bt+1$ is a generic zero of $f$,
where $a$ and $b$ are arbitrary constants.
So ${\rm in}_{{S}}(I)$ has no monomials.
But ${\rm in}_{{S}}(f)=3y-3$ implies $y'\in[{\rm in}_{{S}}(f):f\in I]$.
\end{exmp}

The following example shows that  a tropical  basis of $I$ may not a generating basis for $I.$
%
\begin{exmp}
Let $S=\{1,3,5,7,\ldots\}$ and $I=[y]$. Then ${\rm in}_S(I)=(y,y',y'',\cdots)$.
Let $g=y+ty'.$
Then $\delta^{2i}g=(2i+1)\delta^{2i}y+t\delta^{2i+1}y$ and ${\rm in}_S(\delta^{2i}g)=\delta^{2i}g$,
$\delta^{2i+1}g=(2i+2)\delta^{2i+1}y+t\delta^{2i+2}y$ and ${\rm in}_S(\delta^{2i+1}g)=\delta^{2i+1}y$ for $i\in\mathbb{N}.$
So $({\rm in}_S(\delta^jg):j\in\mathbb{N})=(y,y',y'',\cdots)={\rm in}_S(I)$, which implies that $\{g\}$ is a tropical basis of $I$ but $I\neq [g].$
\end{exmp}

The following example shows that a linear differential polynomial $f$ may not be a tropical basis of $[f]$ and the tropical  basis of $[f]$ may consist more than one polynomials.
\begin{exmp}\label{exmp2.6}
Let $S=4\mathbb{N}$, $f=\delta^4y+y''+y'$, and $I=[f]\subset\F\{y\}.$ Then  for $n\in\mathbb{N},$
\begin{equation}
\left\{
             \begin{array}{lr}
             \delta^{4n}f=\delta^{4n+4}y+\delta^{4n+2}y+\delta^{4n+1}y, &  \\
             \delta^{4n+1}f=\delta^{4n+3}y+\delta^{4n+2}y+\delta^{4n+5}y, &  \\
             \delta^{4n+2}f=\delta^{4n+4}y+\delta^{4n+3}y+\delta^{4n+6}y, &  \\
             \delta^{4n+3}f=\delta^{4n+4}y+\delta^{4n+7}y+\delta^{4n+5}y,  &
             \end{array}
\right.
\end{equation}
and
\begin{equation}
\left\{
             \begin{array}{lr}
             {\rm in}_{S}(\delta^{4n} f)=\delta^{4n+4}y, & \\
             {\rm in}_S(\delta^{4n+1}f)=\delta^{4n+3}y,  & \\
             {\rm in}_S(\delta^{4n+2}f)=\delta^{4n+4}y,  & \\
             {\rm in}_S(\delta^{4n+3}f)=\delta^{4n+4}y,  &
             \end{array}
\right.
\end{equation}
So $[{\rm in}_S(\delta^if):i\in\mathbb{N}]=(\delta^{4(n+1)}y,\delta^{4n+3}:n\in\mathbb{N}).$
But $g=3y''+\delta^9y+2y'=2f+\delta^5f+f'-f''-f'''\in[f]$ and ${\rm in}_S(g)=3y''\notin (\delta^{4(n+1)}y,\delta^{4n+3}:n\in\mathbb{N})$.
In fact, $\{f,g,\delta^6y-2y''-\delta^5y-y',\delta^{13}y-2\delta^9y+5\delta^5-y'\}$ is a tropical differential Gr\"obner basis of $I$ with respect to $S,$ which can be computed by the algorithm Tr-DGB in section 4.
\end{exmp}

\section{Tropical differential Gr\"obner basis}
\label{sec.2}
\subsection{Tropical differential Gr\"obner basis}
\label{sec.2.1}

\begin{defn}
Let $\prec$ be a total ordering on the set $\mathcal{M}_\Y$ of monomials of $\F\{\Y\}$.
The order $\prec$ is said to be an admissible monomial ordering if for $M,N,U\in\mathcal{M}_\Y$
\begin{itemize}
\item $1\prec M$ for $M\neq 1$,
\item $M \prec N$ implies $UM\prec UN$,
\item $M$ $ \prec$ the maximal monomial in $\delta M$, for $M\neq 1,$
\item $M \prec N$ implies the maximal monomial  in $\delta M$ $\prec$ the maximal monomial in $\delta N$.
\end{itemize}
If $\prec$ is admissible,
we denote by $\lm_\prec(P)$ the maximal monomial in $P$, called the {\em leading monomial},
and by $\lc_\prec(P)$ the coefficient of $\lm_\prec(P)$ in $P$, called the {\em leading coefficient}.
Set $\lt_\prec(P)=\lc_\prec(P)\lm_\prec(P)$, which is called the {\em leading term} of $P.$
\end{defn}

In the rest of this section, let ${\overline{S}}\in \mathcal{P}(\mathbb{N})^n$ and $\prec$ be an admissible monomial ordering.
For $f\in \F\{\Y\}$ and ${\overline{S}}\in \mathcal{P}(\mathbb{N})^n$,
if ${\rm lt}_\prec({\rm in}_{{\overline{S}}}(f))=cE_M$,
we denote by $\lm_{{\overline{S}}}(f)=E_M$ and by $\lc_{{\overline{S}}}(f)$ the coefficient of $E_M$ in $f$. Set $\lt_{{\overline{S}}}(f)=\lc_{{\overline{S}}}(f)\lm_{{\overline{S}}}(f)$. Note that $\lc_{{\overline{S}}}(f)\in \F$.

\begin{defn}
{
A subset $G$ of a differential ideal $I$ is called {\em tropical differential Gr\"obner basis} of $I$ if $G$ is a basis of $I$ and $\lm_{{\overline{S}}}(I):=(\lm_{{\overline{S}}}(f):f\in I)=(\lm_{{\overline{S}}}(\delta^ig):i\in\mathbb{N},g\in G)$.}
\end{defn}

From Example \ref{exmp2.6}, we can see that $f$ is a differential Gr\"obner basis of $[f]$,
which may not be a tropical differential Gr\"obner basis.

\begin{lem}
A tropical differential  Gr\"obner basis $G$ of $I$ with respect to ${\overline{S}}$ is a tropical basis with respect to ${\overline{S}}$.
\end{lem}
{\noindent\bf Proof.}
It suffices to show that for any $f\in I$,
${\rm in}_{{\overline{S}}}(f)\in {\rm in}_G = ({\rm in}_{{\overline{S}}}(\delta^k g), g\in G, k\in\N)$.
Since $G$ is a tropical differential  Gr\"obner basis of $I$,
for any  $f\in I$, we have $\lm_{{\overline{S}}}(f)\in (\lm_{{\overline{S}}}(\delta^ig):i\in\mathbb{N},g\in G)$.
Then there exists an $i\in\mathbb{N}$, $g\in G$ and a term $T$ such that $\lm_{{\overline{S}}}(f)=T\lm_{{\overline{S}}}(\delta^ig)$.
So $h=f-T\delta^ig\in I$. If ${\rm in}_{{\overline{S}}}(f)-T{\rm in}_{{\overline{S}}}(\delta^ig)= 0$, we already have ${\rm in}_{{\overline{S}}}(f)\in {\rm in}_G$.
Suppose ${\rm in}_{{\overline{S}}}(f)-T{\rm in}_{{\overline{S}}}(\delta^ig)\neq 0$.
Then ${\rm in}_{{\overline{S}}}(h)={\rm in}_{{\overline{S}}}(f)-T{\rm in}_{{\overline{S}}}(\delta^ig)$ and
the valuation of each term in ${\rm in}_{{\overline{S}}}(h)$ is equal to that of ${\rm in}_{{\overline{S}}}(f)$
and ${\rm in}_{{\overline{S}}}(h)\prec {\rm in}_{{\overline{S}}}(f).$
Repeating the above process to $h$, $ {\rm in}_{{\overline{S}}}(f)=\sum_{i=1}^N T_i{\rm in}_{{\overline{S}}}(g_i)$ for some $g_i\in\{\delta^jg:i\in\mathbb{N},g\in G\}$ and $N\in\mathbb{N}.$ The lemma is proved.
\qed

\begin{defn}
We extend the mapping $\Val_{{\overline{S}}}$ into $\F\{\Y\}$: $\Val_{{\overline{S}}}(f)=\val(\lc_{{\overline{S}}}(f))+\Val_{{\overline{S}}}(\lm_{{\overline{S}}}(f)),$ $f\in\F\{\Y\}.$
\end{defn}

\begin{lem}\label{lm}
For $f,g\in \F\{\Y\}$ and ${\overline{S}}\in \mathcal{P}(\mathbb{N})^n$, $\Val_{{\overline{S}}}(f+g)\geq\min\{\Val_{{\overline{S}}}(f),\Val_{{\overline{S}}}(g)\}.$
\end{lem}
{\noindent\bf Proof.}
Let $f=\lc_{{\overline{S}}}(f)\lm_{{\overline{S}}}(f)+\cdots$ and $g=\lc_{{\overline{S}}}(g)\lm_{{\overline{S}}}(g)+\cdots$.
Then $f+g=\lc_{{\overline{S}}}(f)\lm_{{\overline{S}}}(f)+\lc_{{\overline{S}}}(g)\lm_{{\overline{S}}}(g)+\cdots$.

If $\Val_{{\overline{S}}}(f)=\Val_{{\overline{S}}}(g)$ and $\lm_{{\overline{S}}}(f)=\lm_{{\overline{S}}}(g)$,
then $\val(\lc_{{\overline{S}}}(f))=\val(\lc_{{\overline{S}}}(g))$.
So $\Val_{{\overline{S}}}(f+g)\geq\Val_{{\overline{S}}}(f)=\Val_{{\overline{S}}}(g).$

If $\Val_{{\overline{S}}}(f)=\Val_{{\overline{S}}}(g)$ and $\lm_{{\overline{S}}}(f)\prec\lm_{{\overline{S}}}(g)$,
then $\lc_{{\overline{S}}}(f+g)\lm_{{\overline{S}}}(f+g)=\lc_{{\overline{S}}}(g)\lm_{{\overline{S}}}(g)$.
So $\Val_{{\overline{S}}}(f+g)=\Val_{{\overline{S}}}(g).$

If $\Val_{{\overline{S}}}(f)\neq\Val_{{\overline{S}}}(g)$  and $\lm_{{\overline{S}}}(f)=\lm_{{\overline{S}}}(g)$,
then $\val(\lc_{{\overline{S}}}(f))\neq\val(\lc_{{\overline{S}}}(g))$,
which implies $\lc_{{\overline{S}}}(f+g)=\lc_{{\overline{S}}}(f)+\lc_{{\overline{S}}}(g)$
and $\lm_{{\overline{S}}}(f+g)=\lm_{{\overline{S}}}(f)=\lm_{{\overline{S}}}(g)$.
By the property of $\val$, $\Val_{{\overline{S}}}(f+g)\geq\min\{\Val_{{\overline{S}}}(f),\Val_{{\overline{S}}}(g)\}.$

If $\Val_{{\overline{S}}}(f)\neq\Val_{{\overline{S}}}(g)$  and $\lm_{{\overline{S}}}(f)\neq\lm_{{\overline{S}}}(g)$,
then $\lm_{{\overline{S}}}(f+g)=\lm_{{\overline{S}}}(f)$ and $\val(\lc_{{\overline{S}}}(f+g))\geq \val(\lc_{{\overline{S}}}(f))$
or  $\lm_{{\overline{S}}}(f+g)=\lm_{{\overline{S}}}(g)$ and $\val(\lc_{{\overline{S}}}(f+g))\geq \val(\lc_{{\overline{S}}}(g))$.
So $\Val_{{\overline{S}}}(f+g)\geq\min\{\Val_{{\overline{S}}}(f),\Val_{{\overline{S}}}(g)\}.$
\qed

A differential polynomial $f$ is called {\em homogeneous} if it is a
homogeneous polynomial in the variables $\Theta \Y$.
Note that this is different from differentially homogenous \cite{sdres}.
%

\begin{defn}
Fix homogeneous differential polynomials $f,g\in \F\{\Y\}$, ${\overline{S}}\in \mathcal{P}(\mathbb{N})^n$ and an admissible monomial  ordering $\prec$.
Write $f=\lc_{{\overline{S}}}\lm_{{\overline{S}}}(f)+\cdots$ and $g=\lc_{{\overline{S}}}(g)\lm_{{\overline{S}}}(g)+\cdots$.
Then $f\prec_{{\overline{S}}}g$ if $\Val_{{\overline{S}}}(f)<\Val_{{\overline{S}}}(g)$ or $\Val_{{\overline{S}}}(f)=\Val_{{\overline{S}}}(g)$ and $\lm_{{\overline{S}}}(g)\prec\lm_{{\overline{S}}}(f).$
In addition we set $f\leq 0$ for all nonzero $f$.
\end{defn}

Let $\prec$ be an admissible monomial ordering and ${\overline{S}}\in \mathcal{P}(\mathbb{N})^n$.
Given two homogeneous differential polynomials $f,g\in\F\{\Y\}$,
if there is no monomial in $f$ which is a multiple of $\lm_{{\overline{S}}}(g)$, we say $f$ is {\em ${\overline{S}}$-reduced} w.r.t. $g$.
If there is no monomial in $f$ which is a multiple of $\lm_{{\overline{S}}}(\delta^ig)$, we say $f$ is {\em ${\overline{S}}$-differentially reduced} w.r.t. $g$.
%
The following algorithm computes the ``normal" form of $f$ w.r.t. a set of differential polynomials.

\noindent {\bf Algorithm: Differential Reduction}\\
{\bf Input:} $N\in\N{>0}$, homogeneous differential polynomials $g_1,\ldots,g_t, f$ in $\F\{\Y\}$,
             ${\overline{S}}\in \mathcal{P}(\mathbb{N})^n$ and an admissible monomial ordering $\prec$.\\
{\bf Output:} Fail or  differential polynomials $h_{ij},r\in \F\{\Y\}$ satisfying
$$f=\sum_{i=1}^t\sum_{k=0}^{N}h_{ik}\delta^kg_i+r,$$
where $f\preceq_{{\overline{S}}}h_{ij}\delta^jg_i$, for $j\in\mathbb{N}$, $1\leq i\leq n$, and $f\preceq_{{\overline{S}}}r$.
Besides, $r$ is ${\overline{S}}$-differentially reduced w.r.t. $g_1,\ldots,g_t$.
\begin{description}
\item{1.} Let $h_{ik}=0$, $i=1,\ldots,t$ and $k\in\N$,  $r=f$, $k=0$.

\item{2.} While $k\le N$ and $r$ is not ${\overline{S}}$-differentially reduced w.r.t. $g_1,\ldots,g_t$
\begin{description}
\item{2.1}  Let $j\in\N$ such that $\lm_{{\overline{S}}}(\delta^jg_i)$ divides $\lm_{{\overline{S}}}(r)$.

\item{2.2} Let $p_{ij}=\lt_{{\overline{S}}}(r)/\lt_{{\overline{S}}}(\delta^jg_i)$, $h_{ij}=h_{ij}+p_{ij}$
   and $r=r - p_{ij}\delta^jg_i$.
\item{2.3} $k=k+1$.
\end{description}
\item{3.} Output the nonzero $h_{ij}$ and $r.$
\end{description}
The output $r$ of the above algorithm is also called the {\em ${\overline{S}}$-differential normal form} of $f$ w.r.t. $G=\{g_1,\ldots,g_t\}$ and we denote $f\xrightarrow[{\rm diff}{\overline{S}}]{G} r.$

In \cite{JA}, in order to make the reduction process terminating, the polynomials obtained in previous procedure can be used to reduce the current one. But in differential case,
the reduction procedure may not terminate even adopting the strategy.
So, we force to algorithm to stop after $N$ steps.

\begin{exmp}
{
Let $S=\mathbb{N}$ and $\prec$ be an admissible monomial ordering in $\mathbb{C}(t)\{y\}.$
$f=y+t^2y'$ and $g=y+ty'$ are two differential polynomials in $\mathbb{C}(t)\{y\}$.
Use $\{g,\delta g,\ldots\}$ to reduce $f$ and we have
$f\xrightarrow[{\overline{S}}]{g} (t-1)ty'\xrightarrow[{\overline{S}}]{\delta g} -\frac{1}{2}(t-1)t^2y''\xrightarrow[{\overline{S}}]{\delta^2 g} \frac{1}{6}(t-1)t^3y'''\xrightarrow[{\overline{S}}]{\delta^3 g}\cdots.$
We can see that $\frac{1}{i!}(t-1)t^iy^{(i)}$ can not be used to reduce $\frac{1}{m!}(t-1)t^my^{(m)}$ for $i<m$ and this reduction does not terminate.}
\end{exmp}

Let ${\overline{S}}=(S_1,\ldots,S_n)=(\{s_{11},s_{12},\dots\},\ldots,\{s_{n1},s_{n2},\dots\})\in \mathcal{P}(\mathbb{N})^n$. Denote $$\gap({\overline{S}})=\max\{s_{ij+1}-s_{ij}:i=1,\ldots,n,j\in\mathbb{N}\}$$ and we call $\gap({\overline{S}})$ the {\em gap} of ${\overline{S}}.$
We show that the Differential Reduction algorithm terminates for homogeneous differential polynomials
if their coefficients are constant and the gap of ${\overline{S}}$ is bounded.

\begin{lem}\label{lm2.15}
Let the gap of  ${\overline{S}}$  be  bounded by $N_0$.
For homogeneous differential polynomials $G$ and $f$
with constant coefficients, the algorithm Differential Reduction terminates.
\end{lem}
{\noindent\bf Proof.}
Since the gap of ${\overline{S}}$ is bounded by $N_0$, for any $\delta^iy_j$, $\Val_{S_j}(\delta^iy_j)\leq N_0$.
So for a homogeneous differential polynomial $f$ of degree $d$ in $\CC\{\Y\}$, $\Val_{{\overline{S}}}(f)\leq dN_0$.
Since the degree of the differential polynomial $r$ in the reduction process
has degree at most $d=\deg(f)$,
the valuations of all $r$ in the process are bounded by $dN_0.$
Denote the polynomial $r$ in loop $i$ by $r_i$.
Since the coefficients are constants,
if the algorithm does not terminate, then there exists an $c\in\mathbb{N}$
such that  $\Val_{{\overline{S}}}(r_c)=\Val_{{\overline{S}}}(r_{c+1})=\ldots$ and
$\lm_{{\overline{S}}}(r_c)\succ\lm_{{\overline{S}}}(r_{c+1})\succ\cdots$, which contradicts to the property of admissible monomial ordering.
\qed

%
\begin{lem}\label{lmm}
Let $G$ be a tropical differential  Gr\"obner basis of $I$ with respect to ${\overline{S}}\in \mathcal{P}(\mathbb{N})^n$. Then for $f\in I$, we have $f\xrightarrow[{\rm diff}{\overline{S}}]{G} 0.$
\end{lem}
{\noindent\bf Proof.}
If the algorithm terminates and $f\xrightarrow[{\rm diff}{\overline{S}}]{G} r\neq 0$, then $r\in I$. So there exists $g\in G$ and $i\in\mathbb{N}$ such that $\lm_{{\overline{S}}}(r)$ is a multiple of $\lm_{{\overline{S}}}(\delta^ig)$, which contradicts the algorithm.
\qed

We can study tropical differential Gr\"obner basis of a differential ideal from the point view of algebra.

Let $\F[\Y^{(o)}]=\F[y_1,\ldots,y_n,\delta y_1,\ldots,\delta y_n,\ldots,\delta^o y_1,\ldots,\delta^o y_n].$
Given ${\overline{S}}=(S_1,\ldots,S_n)\in \mathcal{P}(\mathbb{N})^n$,
set
\begin{eqnarray*}
{w^{(o)}} &=&(
\Val_{S_1}(y_1),\ldots,\Val_{S_n}(y_n),
\Val_{S_1}(\delta y_1),\ldots,\Val_{S_n}(\delta y_n),\\
&&\ldots,\Val_{S_1}(\delta^oy_1),\ldots,\Val_{S_n}(\delta^oy_n)),
\end{eqnarray*}
which is a vector in $\mathbb{N}^{(o+1)n}.$
Note that ${\rm in}_{{\overline{S}}}(f)={\rm in}_{{w^{(o)}}}(f)$ for $f\in \F\{\Y\}\bigcap \F[\Y^{(o)}]$.

Let $G=\{g_1,\ldots,g_m\}\subseteq \F\{\Y\}$ be a set of homogeneous differential polynomials, ${\overline{S}} \in \mathcal{P}(\mathbb{N})^n$ and $\prec$ be
  an admissible monomial ordering.
Set $G^{(o)}=\{g_1,\ldots,g_m,\ldots,\delta^og_1,\ldots,\delta^og_m\}$ for $o\in\mathbb{N}$ and $G^{\infty}=\bigcup_{i=0}^\infty G^{(i)}.$
Then there exists the smallest number $l\in \mathbb{N}$ such that we can regard $G^{(o)}$ as a set of polynomials in the ring $\F[\Y^{(l)}]$.
Then we can use the method in \cite{JA} to compute a Gr\"obner basis $\mathcal{G}^{(o)}$ for the ideal $(G^{(o)})$ in $\F[\Y^{(l)}]$ with respect to ${w^{(l)}}$.

\begin{prop}\label{prop3.9}
Using the above notation, $\mathcal{G}=\bigcup_{i=0}^\infty \mathcal{G}^{(i)}$ is a tropical { differential Gr\"obner}
basis for $[G]$ with respect to ${\overline{S}}$.
\end{prop}
{\noindent\bf Proof.}
First, $G$ is a basis of $[G]$.
For $f\in[G]$, there exist $N,M\in\mathbb{N}$ such that $f\in(G^{(N)})\subseteq\F[\Y^{(M)}]$.
Then by the proof of Algorithm 2.9 in \cite{JA}, ${\rm lm}_{{w^{(M)}}}(f)\in ({\rm lm}_{{w^{(M)}}}(g):g\in \mathcal{G}^{(N)})$.
So ${\rm lm}_{{\overline{S}}}(f)={\rm lm}_{{w^{(M)}}}(f)\in ({\rm lm}_{{w^{(M)}}}(g):g\in \mathcal{G}^{(N)})=({\rm lm}_{{\overline{S}}}(g):g\in \mathcal{G}^{(N)})\subseteq({\rm lm}_{{\overline{S}}}(\delta^ig):g\in\mathcal{G},i\in\mathbb{N})$.
\qed

\subsection{Buchberger style criterion for tropical differential Gr\"obner basis}
\label{sec.3.2}

As in the algebraic case, we can use the normal form algorithm to compute a Gr\"obner basis using the Buchberger algorithm.
Let ${\overline{S}}\in \mathcal{P}(\mathbb{N})^n$ and  $f$ and $g$ be two differential polynomials in $\F\{\Y\}$. We define the $\tr-$S-polynomial of $f$ and $g$ with respect to ${\overline{S}}$ to be
$$ \tr-S(f,g)=\lc_{{\overline{S}}}(g)\frac{\lcm(\lm_{{\overline{S}}}(f),\lm_{{\overline{S}}}(g))}{\lm_{{\overline{S}}}(f)}f-\lc_{{\overline{S}}}(f)\frac{\lcm(\lm_{{\overline{S}}}(f),\lm_{{\overline{S}}}(g))}{\lm_{{\overline{S}}}(g)}g.$$
Then we can use the $\tr-$S-polynomial to obtain the tropical differential Gr\"obner basis. The difference of the proof of following theorem with that of algebraic case is that the indeterminates is infinite.

For $a,b\in\mathbb{N},$ denote $\lfloor a,b \rfloor$ to be $\{a,a+1,\ldots,b-1,b\}.$

\begin{thm}\label{thm1}
Let $G=\{g_1,\ldots,g_t\}$ be a set of homogeneous differential polynomials in $\F\{\Y\}$, ${\overline{S}} \in \mathcal{P}(\mathbb{N})^n$.
If $\tr-S(\delta^l g_i,\delta^k g_j)$ is differentially reduced to zero by $G$ for any $i,j\in \lfloor1,t\rfloor$ and $l,k\in \mathbb{N}$,
then $G$ is a tropical differential Gr\"obner basis of $[G]$ with respect to ${\overline{S}}.$
\end{thm}
{\noindent\bf Proof.}
We  prove that $\lm_{{\overline{S}}}(f)\in (\lm_{{\overline{S}}}(\delta^ig_j)):j=1,\ldots,t,i\in\mathbb{N})$ for any homogeneous differential polynomial $f\in [G]$.
If on the contrary, $f\in [G]$ is a homogenous differential polynomial in $\F\{\Y\}$ with $\lm_{{\overline{S}}}(f)\notin (\lm_{{\overline{S}}}(\delta^ig_j):j=1,\ldots,t,i\in\mathbb{N})$.
We can write $f=\sum_{i=1}^t\sum_{j=0}^N a_{ij}\delta^jg_i$ for some $N\in\mathbb{N}$,
for some homogenous differential polynomials $a_{ij}$.  By Lemma \ref{lm}, $\Val_{{\overline{S}}}(f)\geq \min\{\Val_{{\overline{S}}}(a_{ij}\delta^jg_i)\}$,
then we can assume that  $\min\{\Val_{{\overline{S}}}(a_{ij}\delta^jg_i)\}$ is maximal over representation of $f$.
For simplicity, we write $f=\sum_{i=1}^t\sum_{j=0}^N a_{ij}\delta^jg_i=\sum_{i=1}^kb_if_i$, where $f_i$ is some $\delta^jg_i$ and $b_k$ is corresponded to $a_{ij}$.
After renumbering we may assume that $\min(\Val_{{\overline{S}}}(b_if_i)))=\Val_{{\overline{S}}}(b_if_i))$ for $1\leqslant i\leqslant d$
and in addition $\lm_{{\overline{S}}}(b_if_i)=\lm_{{\overline{S}}}(b_1f_1)$ for $1\leqslant i\leqslant d'\leqslant d$
with $\lm_{{\overline{S}}}(b_1f_1)$ the largest $\lm_{{\overline{S}}}(b_if_i)$ for $i\leqslant d$.
We may further assume that $d'$ is as small as possible among descriptions achieving the maximum.
Since $\lm_{{\overline{S}}}(b_if_i)=\lm_{{\overline{S}}}(b_i)\lm_{{\overline{S}}}(f_i)\in (\lm_{{\overline{S}}}(\delta^ig_j):j=1,\ldots,t,i\in\mathbb{N})$, $\val(\lc_{{\overline{S}}}(b_1f_1+\ldots+b_{d'}f_{d'}))>\min(\val(\lc_{{\overline{S}}}(b_if_i)))$ for $i=1,\ldots,d'.$
Thus $d'\geq 2.$ Because in the representation of $f$, $N$ can be choose as big as possible, we can write $\tr-S(f_1,f_2)=\sum_{i=1}^kh_if_i$ with $h_if_i\succeq \tr-S(f_1,f_2)$ by hypothesis.
Then
$$
\begin{aligned}
f&=\sum_{i=1}^kb_if_i
\\&=\sum_{i=1}^kb_if_i-\frac{\lc_{{\overline{S}}}(b_1f_1)\lm_{{\overline{S}}}(b_1f_1)}{\lc_{{\overline{S}}}(f_1)\lc_{{\overline{S}}}(f_2)\lcm(\lm_{{\overline{S}}}(f_1),\lm_{{\overline{S}}}(f_2))}(\tr-S(f_1,f_2)-\sum_{i=1}^kh_if_i)\\
&=(b_1-\frac{\lc_{{\overline{S}}}(b_1f_1)\lm_{{\overline{S}}}(b_1f_1)}{\lc_{{\overline{S}}}(f_1)\lm_{{\overline{S}}}(f_1)}
+\frac{\lc_{{\overline{S}}}(b_1f_1)\lm_{{\overline{S}}}(b_1f_1)}{\lc_{{\overline{S}}}(f_1)\lc_{{\overline{S}}}(f_2)\lcm(\lm_{{\overline{S}}}(f_1),\lm_{{\overline{S}}}(f_2))}h_1)f_1
\\&+(b_2+\frac{\lc_{{\overline{S}}}(b_1f_1)\lm_{{\overline{S}}}(b_1f_1)}{\lc_{{\overline{S}}}(f_2)\lm_{{\overline{S}}}(f_2)}
+\frac{\lc_{{\overline{S}}}(b_1f_1)\lm_{{\overline{S}}}(b_1f_1)}{\lc_{{\overline{S}}}(f_1)\lc_{{\overline{S}}}(f_2)\lcm(\lm_{{\overline{S}}}(f_1),\lm_{{\overline{S}}}(f_2))}h_2)f_2
\\&+\sum_{i=3}^k(b_i+\frac{\lc_{{\overline{S}}}(b_1f_1)\lm_{{\overline{S}}}(b_1f_1)}{\lc_{{\overline{S}}}(f_1)\lc_{{\overline{S}}}(f_2)\lcm(\lm_{{\overline{S}}}(f_1),\lm_{{\overline{S}}}(f_2))}h_i)f_i
\\&=\sum_{i=1}^k\widetilde{b_i}f_i,
\end{aligned}
$$
where $\widetilde{b_i}$ is defined to be the polynomial multiplying $f_i$ in the previous line.
By construction $\widetilde{b_1}\succ b_1$ and $\widetilde{b_i}\succ b_i$ for all $i\geq 2.$
Thus we have a new representation of $f$ with either $\min\{\Val_{{\overline{S}}}(\widetilde{b_i}f_i)\}$ larger or this minimum the same and $d'$ smaller,
which contradicts our assumptions on the respective maximality and minimality of these quantities. So we prove the claim.
\qed

Let $f,g\in \F\{\Y\}$. If $\gcd(\lm_{{\overline{S}}}(f),\lm_{{\overline{S}}}(g))=1$,
then we can give the following lemma which is similar to the algebraic Gr\"obner basis.
\begin{lem}\label{lm1}
If
$f,g\in\F\{\Y\}$ satisfy $\gcd(\lm_{{\overline{S}}}(f),\lm_{{\overline{S}}}(g))=1$,
then $\tr-S(f,g)$ $\xrightarrow[{\overline{S}}]{\{f,g\}} 0$.
\end{lem}
{\noindent\bf Proof.}
Assume on the contrary that $\tr-S(f,g)\xrightarrow[{\overline{S}}]{\{f,g\}} h\neq 0.$ Then $h=pf-qg$ for some $p,q\in\F\{\Y\}.$
By Lemma \ref{lm}, $\val_{{\overline{S}}}(h)\geq\min\{\val_{{\overline{S}}}(pf),\val_{{\overline{S}}}(qg)\}.$
If $\val_{{\overline{S}}}(h)>\min\{\val_{{\overline{S}}}(pf),\val_{{\overline{S}}}(qg)\},$ then $\val_{{\overline{S}}}(pf)=\val_{{\overline{S}}}(qg)$ and $\lm_{{\overline{S}}}(pf)=\lm_{{\overline{S}}}(qg).$
Because $\gcd(\lm_{{\overline{S}}}(f),\lm_{{\overline{S}}}(g))=1$, there is $M\in\mathcal{M}$ such that
$\lm_{{\overline{S}}}(p)=M\lm_{{\overline{S}}}(g)$ and $\lm_{{\overline{S}}}(q)=M\lm_{{\overline{S}}}(f)$.
Let $\bar{p}=p-\frac{\coeff(\lc_{{\overline{S}}}(p))}{\coeff(\lc_{{\overline{S}}}(g))}Mg$ and $\bar{q}=q-\frac{\coeff(\lc_{{\overline{S}}}(q))}{\coeff(\lc_{{\overline{S}}}(f))}Mf$.
Then $h=\bar{p}f-\bar{q}g$ and $\val_{{\overline{S}}}(\bar{p}f)>\val_{{\overline{S}}}(pf)$, $\val_{{\overline{S}}}(\bar{q}g)>\val_{{\overline{S}}}(qg).$
If $\val_{{\overline{S}}}(h)>\min\{\val_{{\overline{S}}}(\bar{p}f),\val_{{\overline{S}}}(\bar{q}g)\},$
then repeating the above process and obtain an expression for $h$ with $h=p'f-q'g$ and $\val_{{\overline{S}}}(h)=\min\{\val_{{\overline{S}}}(p'f),\val_{{\overline{S}}}(q'g)\}.$

Without loss of generality, we assume that $\val_{{\overline{S}}}(h)=\val_{{\overline{S}}}(p'f)\leq \val_{{\overline{S}}}(q'g)$.
Suppose $\lm_{{\overline{S}}}(h)\prec\lm_{{\overline{S}}}(p'f)$.
Then $\lt_{{\overline{S}}}(p'f)=\lt_{{\overline{S}}}(q'g)$,
which implies $\val_{{\overline{S}}}(p'f)= \val_{{\overline{S}}}(q'g)$.
Since $\gcd(\lm_{{\overline{S}}}(f),\lm_{{\overline{S}}}(g))=1$,
one see that there exists $u\in\mathcal{M}$ such that $\lm_{{\overline{S}}}(p')=u\lm_{{\overline{S}}}(g)$ and $\lm_{{\overline{S}}}(q')=u\lm_{{\overline{S}}}(f)$.
Let $\bar{p'}=p'-\lc_{{\overline{S}}}(p')/\lc_{{\overline{S}}}(g)ug$ and $\bar{q'}=q'-\lc_{{\overline{S}}}(q')/\lc_{{\overline{S}}}(f)uf$.
Then $h=\bar{p'}f-\bar{q'}g$.
We can see that $\val_{{\overline{S}}}(h)=\min\{\val_{{\overline{S}}}(\bar{p'}f),\val_{{\overline{S}}}(\bar{q'}g)\}$,
$\bar{p'}\prec p'$ and $\bar{q'}\prec q'$.
If $\lm_{{\overline{S}}}(\bar{p'}f)$ or $\lm_{{\overline{S}}}(\bar{q'}g)$ is higher than $\lm_{{\overline{S}}}(h)$,
then we repeat the above process and obtain an expression for $h$ with $h=\tilde{p}f-\tilde{q}g$,
where $\lm_{{\overline{S}}}(h)$ is equal to either $\lm_{{\overline{S}}}(\tilde{p}f)$ or $\lm_{{\overline{S}}}(\tilde{q}g).$
This implies that $\lm_{{\overline{S}}}(h)$ is a multiple of $\lm_{{\overline{S}}}(f)$ or $\lm_{{\overline{S}}}(g)$,
which is a contradiction to the assumption that $h$ is ${\overline{S}}$-reduced w.r.t. $\{f,g\}.$ Hence $h=0.$
\qed

\subsection{Lower bound for differentiation and a possible Buchberger style algorithm}
\label{sec.3.21}

Theorem \ref{thm1} is not an effective criterion, since we need to check whether  
$\tr-S(\delta^l g_i,\delta^k g_j)\xrightarrow[{\rm diff}{\overline{S}}]{G} 0$ holds for all $l,k\in \mathbb{N}$ and $g_i,g_j\in G.$
In the rest of this section, we give a lower bound for number of differentiations $l$ and $k$ needed to compute a Gr\"obner basis.

For ${\overline{S}}=\mathbb{N}^n$ and $f\in \CC[t]\{\Y\}$,   we can write $f$ as
$f=f_0+tf_1+\cdots+t^df_d,$
for some $d\in\mathbb{N}$ and $f_0,f_1,\ldots,f_d\in \CC\{\Y\}$.
Then for $0\leq k\leq d,$
\begin{equation}\label{equ-8}
\begin{aligned}
\delta^kf&=\delta^kf_0+\delta^k(tf_1)+\cdots+\delta^k(t^df_d)
\\&=(\delta^kf_0+b_{1,k-1}\delta^{k-1}f_1+\cdots+b_{k,0}f_k)
\\&+(b_{1,k}\delta^kf_1+b_{2,k-1}\delta^{k-1}f_2+\cdots+b_{k,1}f_k'+b_{k+1,0}f_{k+1})t
\\&+\cdots
\\&+(b_{l-k,k}\delta^kf_{l-k}+b_{l-k+1,k-1}\delta^{k-1}f_{l-k+1}+\cdots+b_{l,0}f_l)t^{l-k}
\\&+(b_{d-k+1,k}\delta^kf_{d-k+1}+b_{d-k+2,k-1}\delta^{k-1}f_{d-k+2}+\cdots+b_{d,1}f_d')t^{d-k+1}
\\&+\cdots
\\&+(b_{d-1,k}\delta^kf_{d-1}+b_{d,k-1}\delta^{k-1}f_d)t^{d-1}
\\&+b_{d,k}\delta^kf_dt^d,
\end{aligned}
\end{equation}
for some  $b_{i,j}\in \CC.$
For $k> d$,
\begin{equation}\label{equ-9}
\begin{aligned}
\delta^kf&=\delta^kf_0
\\&+(b_{1,k}t\delta^kf_1+b_{1,k-1}\delta^{k-1}f_1)
\\&+(b_{2,k}t^2\delta^kf_2+b_{2,k-1}t^{1}\delta^{k-1}f_2+b_{k,k-2}\delta^{k-2}f_2)
\\&+\cdots
\\&+(b_{d,k}t^d\delta^kf_d+b_{k,k-1}t^{d-1}\delta^{k-1}f_d+\cdots+b_{d,1}t\delta f_d+b_{d,k-d}t^{k-d}f_d)
\\&=(\delta^kf_0+b_{1,k-1}\delta^{k-1}f_1+\cdots+b_{d,k-d}f_d)
\\&+(b_{1,k}\delta^kf_1+b_{2,k-1}\delta^{k-1}f_2+\cdots+b_{d-1,k-d+2}f_{d-1}'+b_{d,k-d+1}f_{d})t
\\&+\cdots
\\&+(b_{l,k}\delta^kf_{l}+b_{l+1,k-1}\delta^{k-1}f_{l+1}+\cdots+b_{d,k+l-d}f_d)t^{l}
\\&+\cdots
\\&+b_{d,k}\delta^kf_dt^d,
\end{aligned}
\end{equation}
for some $b_{i,j}\in \CC.$

If $\delta^kf_0+b_{1,k-1}\delta^{k-1}f_1+\cdots+b_{k,0}f_k$ in equation (\ref{equ-8})
and $\delta^kf_0+b_{1,k-1}\delta^{k-1}f_1+\cdots+b_{d,k-d}f_d$ in equation (\ref{equ-9}) are not zero, then
\begin{equation}\label{equ-10}
{\rm in}_{{\overline{S}}}(\delta^kf)=\left\{
\begin{array}{rcl}
\delta^kf_0+b_{1,k-1}\delta^{k-1}f_1+\cdots+b_{k,0}f_k,    &         & {0\leq k\leq d}\\
\delta^kf_0+b_{1,k-1}\delta^{k-1}f_1+\cdots+b_{d,k-d}f_d,          &         & {k> d}.
\end{array}\right.
\end{equation}

Based on (\ref{equ-10}), we also have the following simple fact.
\begin{lem}
Let ${\overline{S}}=\mathbb{N}^n$, $\prec$ be an admissible monomial ordering satisfying $\delta^{j_1}y_{i_1}\prec\delta^{j_2}y_{i_2}$
if $j_1<j_2$ for $j_1,j_2\in\mathbb{N}$ and $i_1,i_2\in\lfloor1,m\rfloor$ and $f$ be a linear homogeneous differential polynomial in $\CC[t]\{\Y\}$.
Write $f$ as $f=f_0+tf_1+\cdots+t^df_d$.
If $\ord(f_0)>\ord(f_i)-i$ for $i\geq 1$,
then $\{f\}$ is a tropical differential Gr\"obner basis of $[f]$.
\end{lem}
{\noindent\bf Proof.}
$\ord(f_0)>\ord(f_i)-i$ implies that
$\delta^kf_0+b_{1,k-1}\delta^{k-1}f_1+\cdots+b_{k,0}f_k$ in equation (\ref{equ-8}) and $\delta^kf_0+b_{1,k-1}\delta^{k-1}f_1+\cdots+b_{d,k-d}f_d$ in equation (\ref{equ-9}) are not zero.
So equation (\ref{equ-10}) holds. Therefore $\lm_{{\overline{S}}}(\delta^kf)\neq \lm_{{\overline{S}}}(\delta^lf)$ for any $k,l\in\mathbb{N}$.
By Theorem \ref{thm1} and Lemma \ref{lm1}, the lemma is proved.
\qed

From (\ref{equ-10}),  $f_d$ occurs in ${\rm in}_{{\overline{S}}}(\delta^kf)$
for $k\ge d$. Thus, $d=\deg(f,t)$ is a lower bound of the number of differentiations
for $f$ to compute the tropical differential Gr\"obner basis of $[f]$.
It is easy to show that  $n$ and $\gap(\overline{S})$ are also such lower bounds
in certain cases.

We give another lower bound related with the orders of  $f$.
For $f\in\F\{\Y\}$, denote $\underline{o}_f$ to be the minimal number such that
 $\delta^{\underline{o}_f}y_i$ appears in $f$ for some $i$.

For $S=2\N$ and $f=y+\delta^5y$.
We differentiate $f$ and obtain
\begin{equation}
\left\{
             \begin{array}{lr}
             f'=\delta^6y+y', & \\
             f''=y''+\delta^7y,  & \\
             \delta^3f=\delta^8y+\delta^3y,  & \\
             \delta^4f=\delta^4y+\delta^9y,  &  \\
             \delta^5f=\delta^{10}y+\delta^5y, &\\
             \delta^6f=\delta^6y+\delta^{11}y. &
             \end{array}
\right.
\end{equation}
By lemma \ref{lm1}, $\tr-S(\delta^if,\delta^jg)$ is not trivial only for $i=0$ and $j=6$, which means that we need to differentiate $f$ $\ord(f)-\underline{o}_f+1$ times in order to compute the tropical differential Gr\"obner basis. This example motivates us to give another lower bound for the number of differentials: $\ord(f)-\underline{o}_f+2\gap(S)$, which will be used in section 4. It is easy to see that the sum of the two lower bounds $\deg(f,t)+\ord(f)-\underline{o}_f+2\gap(S)$ is also a lower bound.

Based on above observation, we
propose the following possible Buchberger style algorithm to compute the tropical Gr\"obner basis for $[G]$.

\noindent {\bf Algorithm: Tropical Gr\"obner Basis}\\
{\bf Input:} $N\in\N_{>0}$, $G=\{g_1,\ldots,g_t\}\subset\C[t]\{\Y\}$, ${\overline{S}} \in \mathcal{P}(\mathbb{N})^n$ and an admissible monomial ordering $\prec$.\\
{\bf Output:} Fail or a tropical Gr\"obner basis of $[G]$.
\begin{description}
\item{1.} Let $d = \max\{n, \max_l\{\deg(g_l,t)+\max_i\{\ord(g_i)\}-\min_j\{\underline{o}_{g_j}\}+2\gap(\overline{S})\}$ and
$G_0 = GB_0= G$, $k=1$.

\item{2.} While $k\le N$
\begin{description}
\item{2.1}  Let $G_k = G_{k-1}\cup
     \{\delta^{(k-1)d+i} g_j\,|\, j=1,\ldots,t, i =1,2,\ldots, d\}$.

\item{2.2} Compute the tropical algebraic Gr\"obner basis $GB_k$ of $G_k$ with the method in \cite{JA}.

\item{2.3} Let $R_k=\{r\neq 0\,|\, f\xrightarrow[{\rm diff}{\overline{S}}]{GB_{k-1}} r\hbox{ for } f\in GB_k\}.$

\item{2.4} If $R_k=\emptyset$ then return $GB_k$; else $k=k+1$.
\end{description}
\item{3.} Fail.
\end{description}

The idea of the algorithm is that in each loop in step 2, we differentiate the differential polynomials in $G$ for an extra $d$ times and terminate the algorithm if we obtain the "same" results in two consecutive loops. We will show in the next section that the algorithm in correct if $G$ consists of homogeneous linear differential polynomials with constant coefficients. In the general case, we conjecture that the algorithm is correct if $[G]$ has a finite differential tropical Gr\"obner basis.

\section{Tropical differential Gr\"obner basis of linear system with constant coefficients}
\label{sec.4}

In this section, we give a complete algorithm for linear homogeneous differential polynomial systems with constant coefficients.

\noindent {\bf Algorithm: Tr-DGB}\\
{\bf Input:} A set of linear homogeneous differential polynomials $\{f_1,\ldots, f_p\}\subseteq \CC\{\Y\}$,
${\overline{S}}=(l_1+m_1\mathbb{N},l_2+m_2\mathbb{N},\ldots,l_n+m_n\mathbb{N})\in\mathcal{P}(\mathbb{N})^n$ and an admissible monomial ordering $\prec$, where $l_1,\cdots,l_n,m_1,$ $\cdots,m_n\in\mathbb{N}$.\\
{\bf Output:}  An integer $k$ and a tropical differential Gr\"obner basis $G=\{g_1,\ldots,g_k\}$ of $I=[f_1,\ldots,f_p]$
satisfying ${\rm in}_S(I)=({\rm in}_{{\overline{S}}}(\delta^jg_i):i=1,\ldots,k,j\in\mathbb{N}).$
\begin{description}
\item[1.] $k=p,$ $c=1$, $L=\lcm(m_1,\ldots,m_n)$, $G=\{f_1,\ldots,f_p\}$.
\item[2.]  $M=\max\{\ord(h_1)-\underline{o}_{h_2}:h_1,h_2\in G\}=(q_k-1)L+r_k$ with $q_k\geq 1$, $r_k<L$.
\item[3.] $\mathcal{S}=\{(\delta^ih_1,\delta^jh_2):h_1,h_2\in G,i,j\in\lfloor0,(q_k+1)L-1\rfloor\}$.
\item[4.] While $\mathcal{S}\neq \varnothing$
\begin{description}
\item[4.1] Pick $\mathcal{T}_c=(a_1,a_2)\in\mathcal{S}$.

\item[4.2] Let $b_k$ be the ${\overline{S}}$-differential normal form of $\tr-S(a_1,a_2)$ w.r.t. $G$.
\item[4.3] If $b_k=0$, then $\mathcal{S}=\mathcal{S}\backslash\{\mathcal{T}_{c}\}$ and $c=c+1.$
\item[4.4] If $b_k\neq 0$, then

   set $G=G\bigcup\{b_k\}$,
    $k=k+1$,

    $M=\max\{\ord(h_1)-\underline{o}_{h_2}:h_1,h_2\in G\}=(q_k-1)L+r_k$ with $q_k\geq 1$,  $r_k<L$,

    $\mathcal{S}=\{(\delta^ih_1,\delta^jh_2):h_1,h_2\in G,i,j\in\lfloor0,(q_k+1)L-1\rfloor\}\backslash\{\mathcal{T}_1,\cdots,\mathcal{T}_{c}\}$, $c=c+1$.

\end{description}
\item[5.] Return $k$ and $G.$
\end{description}

\begin{exmp}\label{exmp4.21}
Let $S=4\mathbb{N}$, $f=\delta^4y+y''+y'$, and $I=[f]\subset\F\{y\}$.
We consider the loop in Step 4 of the algorithm.

Loop 1. $L=4$, $k=1$, $G=\{f\}$, $M=3$, and $q_1=1$. $\mathcal{P}=\{(\delta^if,\delta^jf):i,j\in\lfloor0,7\rfloor\},$ $\tr-S(f,f'')\xrightarrow[{\rm diff}{{\overline{S}}}]{G}-b_1$, where $b_1=\delta^6y-2y''-\delta^5y-y'$ with ${\rm in}_{S}(b_1)=\delta^6y-2y''.$

Loop 2. $L=4$, $k=2$, $G=\{f,b_1\}$, $M=5$, and $q_2=2$. $\mathcal{P}=\{(\delta^if,\delta^jf),$ $(\delta^ib_1,\delta^jb_1),(\delta^if,\delta^jb_1):i,j\in\lfloor0,11\rfloor\}\backslash(f,f'')$,
$\tr-S(f,f''')=f-f'''\xrightarrow[{\rm diff}{{\overline{S}}}]{G} b_2,$ where $b_2=3y''+\delta^9y+2y'$ with ${\rm in}_{S}(b_2)=3y''.$

Loop 3. $L=4$, $k=3,$ $G=\{f,b_1,b_2\}$, $M=8$, and $q_3=3$. $\mathcal{P}=\{(\delta^if,\delta^jf),(\delta^ib_k,\delta^jb_k),(\delta^if,$ $\delta^jb_k):i,j\in\lfloor0,15\rfloor,k=1,2\}
\backslash\{(f,f''),(f,f''')\}$,
$\tr-S(b_1,\delta^4b_2)=3b_1-\delta^4b_2\xrightarrow[{\rm diff}{{\overline{S}}}]{G} -b_3,$ where $b_3=\delta^{13}y-2\delta^9y+5\delta^5y-y'$ with  ${\rm in}_{{S}}(b_3)=b_3.$

Loop 4. $L=4$, $k=4,$ $G=\{f,b_1,b_2,b_3\}$, $M=13$, and $q_4=4$. $\mathcal{P}=\{(\delta^if,\delta^jf),(\delta^ib_k,\delta^jb_k),$ $(\delta^if,\delta^jb_k):i,j\in\lfloor0,19\rfloor,k=1,2,3\}
\backslash\{(f,f''),(f,f'''),$ $(b_1,\delta^4b_2)\}$,
it is easy to verify that $\tr-S(g,h)$ $\xrightarrow[{\rm diff}{{\overline{S}}}]{G} 0$ for $(g,h)\in\mathcal{P}.$

So $G=\{f,b_1,b_2,b_3\}$ is a tropical differential Gr\"obner basis of $[f]$.
\end{exmp}

\begin{thm}\label{thm2}
The Algorithm Tr-DGB  terminates and the output is a tropical differential Gr\"obner basis of the ideal $[f_1,\ldots,f_p]$ with respect to ${\overline{S}}.$
\end{thm}

\begin{lem}\label{lm2}
Fix $m_1,\ldots,m_n,l_1,\cdots,l_n\in\mathbb{N}$.
Let ${\overline{S}}=(l_1+m_1\mathbb{N},\ldots,l_n+m_n\mathbb{N})$.
Denote $L=\lcm(m_1,\ldots,m_n)$ and $L=d_jm_j,j=1,\ldots,n.$
Then $\Val_{{\overline{S}}}(\delta^iy_j)=\Val_{{\overline{S}}}(\delta^{i+kL}y_j)$ for $k,i\in\mathbb{N}$ and $j\in\lfloor1,n\rfloor.$
Moreover, $\lm_{{\overline{S}}}(\delta^{i+kL}f)=\delta^{kL}\lm_{{\overline{S}}}(\delta^{i}f)$ and $\lc_{{\overline{S}}}(\delta^{i+kL}f)=\lc_{{\overline{S}}}(\delta^{i}f)$ for a linear homogeneous differential polynomial $f\in \CC\{\Y\}.$
\end{lem}
{\noindent\bf Proof.}
Assume $i=dm_j+b\in\lfloor dm_j,(d+1)m_j-1\rfloor$, $d,b\in\mathbb{N}$ and  $0\leq b<m_j$.
Then $i+kL=kL+dm_j+b=(d+kd_j)m_j+b\in\lfloor(d+kd_j)m_j,(d+kd_j+1)m_j-1\rfloor.$
By definition of $\Val_{{\overline{S}}}$, we have
$$\Val_{{\overline{S}}}(\delta^{i+kL}y_j)=\Val_{{\overline{S}}}(\delta^iy_j)=\left\{
\begin{array}{rcl}
0,    &         & {b=0}\\
m_j-b,          &         & {b\neq0.}
\end{array}\right.$$
By the definition of admissible monomial ordering,
the lemma is proved.
\qed

\begin{lem}\label{lm4}
Fix $m_1,\ldots,m_n,l_1,\cdots,l_n\in\mathbb{N}$.
Let ${\overline{S}}=(l_1+m_1\mathbb{N},\ldots,l_n+m_n\mathbb{N})$.
Let $G=\{g_1,\ldots,g_p\}\subset \CC\{\Y\}$ be a set of linear homogeneous differential polynomials and $L=\lcm(m_1,\ldots,$ $m_n).$
Set $M=\max\{\ord(g_i)-\underline{o}_{g_j}:i,j\in[1,p]\}=(q_0-1)L+r_0$, where $q_0,r_0\in\mathbb{N}$ and $0\leq r_0\leq L-1.$
If $\tr-S(\delta^l g_i,\delta^k g_j)$ is differentially reduced to zero by $G$ for $i,j\in[1,p]$ and $l,k\in\lfloor0,(q_0+1)L-1\rfloor,$
then $G$ is a tropical differential Gr\"obner basis of $[G]$ with respect to ${\overline{S}}.$
\end{lem}
{\noindent\bf Proof.}
By Theorem \ref{thm1} and Lemma \ref{lm1} , it suffices to show that $\tr-S(\delta^ig_t,\delta^{j}g_k)$ is ${\overline{S}}$-differentially reduced to zero by $\{g_1,\ldots,g_p\},$ for $i,j\in\mathbb{N}$ with $\lm_{{\overline{S}}}(\delta^ig_t)=\lm_{{\overline{S}}}(\delta^jg_k)$.
Suppose $i<j$ and $i=q_1L+r$ with $q_1,r\in\mathbb{N}$ and $0\leq r\leq L-1.$
Then $\ord(g_t)+i\geq \underline{o}_{g_k}+j$ and $\ord(g_k)+j\geq \underline{o}_{g_t}+i$. So $j\leq \ord(g_t)-\underline{o}_{g_k}+i\leq M+i=(q_0-1)L+r_0+q_1L+r<q_1L+(q_0+1)L-1.$
Denote $j=q_1L+d$ with $0\leq d\leq (q_0+1)L-1.$
Then
\begin{eqnarray*}
\tr-S(\delta^ig_t,\delta^{j}g_k)
&=&\lc_{{\overline{S}}}(\delta^jg_k)\delta^ig_t-\lc_{{\overline{S}}}(\delta^ig_t)\delta^jg_k\\
&=&\delta^{q_1L}(\lc_{{\overline{S}}}(\delta^dg_k)\delta^rg_t-\lc_{{\overline{S}}}(\delta^rg_t)\delta^dg_k)\\
&=&\delta^{q_1L}(\tr-S(\delta^rg_t,\delta^{d}g_k))
\end{eqnarray*}
 by Lemma \ref{lm2}.
If $\tr-S(\delta^{r}f,\delta^{d}f)$ is ${\overline{S}}$-differentially reduced to zero by  $\{g_1,\ldots,g_p\}$
and the reduction process is given by
$\tr-S(\delta^{r}f,\delta^{d}f)\xrightarrow[{\overline{S}}]{f_1} h_1\xrightarrow[{\overline{S}}]{f_2} h_2\xrightarrow[{\overline{S}}]{f_3}\cdots\xrightarrow[{\overline{S}}]{f_{l-1}} h_{l-1}\xrightarrow[{\overline{S}}]{f_l} 0$ with $\{f_1,\ldots,f_l\}\subseteq \{g_1,\ldots,g_p,\delta g_1,\ldots,\delta g_p,\ldots\}$ for some $l\in\mathbb{N},$
then $\tr-S(\delta^{i}g_t,\delta^{j}g_k)=\delta^{q_1L}\tr-S(\delta^{r}g_t,\delta^{d}g_k)\xrightarrow[{\overline{S}}]{\delta^{q_1L}f_1} \delta^{q_1L}h_1\xrightarrow[{\overline{S}}]{\delta^{q_1L}f_2} \delta^{q_1L}h_2\xrightarrow[{\overline{S}}]{\delta^{q_1L}f_3}\cdots\xrightarrow[{\overline{S}}]{\delta^{q_1L}f_{l-1}} \delta^{q_1L}h_{l-1}\xrightarrow[{\overline{S}}]{\delta^{q_1L}f_l} 0$ by Lemma \ref{lm2}.
The lemma is proved.
\qed

{\noindent\bf Proof} of Theorem \ref{thm2}:
By Lemma \ref{lm4}, the correctness of Algorithm Tr-DGB is proved.
Next, we prove the termination.
By Lemma \ref{lm2.15}, the process in step 4.2 to compute $b_k$ terminates in finite number of steps.
Then from the algorithm, it is easy to see that $\lm_{{\overline{S}}} (b_i)\neq \lm_{{\overline{S}}} (\delta^l b_j)$ and $\lm_{{\overline{S}}} (b_i)\neq \lm_{{\overline{S}}}(\delta^lf_k)$ for $l\in\mathbb{N}$ and $j<i,k=1,\ldots,p.$
If the algorithm does not terminate,
then there exists an $r\in\lfloor 0,L-1\rfloor$ such that
there exist infinitely many $b_i$ satisfying $\ord(\lm_{{\overline{S}}}(b_i))\mod (L)\equiv r$ and $\lm_{{\overline{S}}}(b_i)\in\{y_l,\delta y_l,\delta^2y_l,\cdots\}$ for some $l\in\lfloor 1,n\rfloor$.
We thus have a set $B=\{b_{i_1},b_{i_2},\ldots\}$ with $i_1<i_2<\ldots$.
If $\ord(\lm_{{\overline{S}}}(b_{i_1}))\leq\ord(\lm_{{\overline{S}}}(b_{i_2}))$, we have $\ord(\lm_{{\overline{S}}}(b_{i_2}))=\ord(\lm_{{\overline{S}}}(b_{i_1}))+dL$ for some $d\in\mathbb{N}$.
Then by Lemma \ref{lm2}, $\lm_{{\overline{S}}}(b_{i_2})=\delta^{dL}\lm_{{\overline{S}}}(b_{i_1})=\lm_{{\overline{S}}}(\delta^{dL}b_{i_1})$, which implies $b_{i_2}$ can be differentially reduced by $b_{i_1}$.
So we have $\ord(\lm_{{\overline{S}}}(b_{i_1}))>\ord(\lm_{{\overline{S}}}(b_{i_2}))>\ldots$, which is impossible.
So the algorithm terminates.
\qed

\begin{defn}
A reduced tropical differential Gr\"obner  basis for an ideal $I\subseteq\F\{\Y\}$
with respect to ${\overline{S}}$ is a tropical differential Gr\"obner  basis for $I$ such that
for all distinct $p,q\in G$, $\lm_{{\overline{S}}}(p)$ is not a multiple of $\lm_{{\overline{S}}}\delta^i q$ for any $i\in\mathbb{N}$.
\end{defn}

\begin{prop}
Fix $m_1,\ldots,m_n,l_1,\cdots,l_n\in\mathbb{N}$.
Let ${\overline{S}}=(l_1+m_1\mathbb{N},l_2+m_2\mathbb{N}$, $\ldots,l_n+m_n\mathbb{N})\in\mathcal{P}(\mathbb{N})^n$.
 Denote $L=\lcm(m_1,\ldots,m_n)$.
$F=\{f_1,\ldots,f_m\}$ is a set of linear homogeneous differential polynomials in $\CC\{\Y\}$.
Then the number of a reduced tropical differential Gr\"obner basis for $[F]$ with respect to ${\overline{S}}$
is not more than $nL.$
\end{prop}
{\noindent\bf Proof.}
Let $G$ be a reduced tropical differential Gr\"obner  basis for $[F]$ with respect to ${\overline{S}}$.
Note that $G$ is a set of linear homogeneous differential polynomials.
If $G$ has more than $nL$ elements, then there exists an $i\in\lfloor1,n\rfloor$ such that $G_0=\{g\in G:\lm_{{\overline{S}}}(g)\in \{y_i,\delta y_i,\cdots\}\}$ has more than $L$ elements.
So there exist
$g_1,g_2\in G_0$ such that
$\ord(\lm_{{\overline{S}}}(g_1))\equiv \ord(\lm_{{\overline{S}}}(g_2))\mod (L)$.
Suppose $\ord(\lm_{{\overline{S}}}(g_1))=\ord(\lm_{{\overline{S}}}(g_2))+kL$ for some $k\in\mathbb{N}.$
Then by Lemma \ref{lm2}, $\lm_{{\overline{S}}}(g_1)=\delta^{kL}\lm_{{\overline{S}}}(g_2)=\lm_{{\overline{S}}}(\delta^{kL}g_2)$, which is a contradiction. The proposition is proved.
\qed

\begin{exmp}
In Example \ref{exmp4.21}, $G=\{f=\delta^4y+y''+y',b_1=\delta^6y-2y''-\delta^5y-y',b_2=3y''+\delta^9y+2y',b_3=\delta^{13}y-2\delta^9y+5\delta^5y-y'\}$ is a tropical differential Gr\"obner basis of $[f]$.
We can verify that $b_1\xrightarrow[{{\overline{S}}}]{\delta^4b_2} -2y''-\frac{5}{3}\delta^5y-\frac{1}{3}\delta^{13}y-y'\xrightarrow[{{\overline{S}}}]{b_2} -\frac{5}{3}\delta^5y-\frac{1}{3}\delta^{13}y+\frac{1}{3}y'+\frac{2}{3}\delta^9y\xrightarrow[{{\overline{S}}}]{b_3} 0.$
So $\{f,b_2,b_3\}$ is a  reduced tropical differential Gr\"obner basis for $[f]$ with respect to ${S}.$
\end{exmp}

\section{Conclusion}

In this paper, we introduced the concept of tropical differential Gr\"obner basis which can be considered as generalizations of the algebraic tropical Gr\"obner basis as well as the tropical differential Gr\"obner basis. We give a Buchberger criterion for tropical  differential Gr\"obner basis and a complete algorithm to compute the tropical differential Gr\"obner basis for a differential ideal generated by homogeneous linear differential polynomials with constant coefficients. In the  general casel, we only give some partial results. We show that the union of the algebraic tropical Gr\"obner basis of $(G^{(d)})$ is a tropical differential  Gr\"obner basis of $[G]$, where $G^{(d)}=\bigcup_{i=0}^d\delta^dG$ and $d\in\N$. We give a lower bound for $d$ such that the algebraic tropical Gr\"obner basis of $(G^{(d)})$ is the differential tropical Gr\"obner basis $[G]$, and based on this observation, we formulated and conjectured a Buchberger style algorithm to compute the differential Gr\"obner basis of $[G]$.

\end{document}